\documentclass[sigconf]{acmart}

\makeatletter                   
\def\mdseries@tt{m}             
\makeatother                    
\usepackage[plain]{fancyref}
\usepackage[draft=true]{minted} 
\usepackage{color}
\usepackage{hyperref}           
\hypersetup{
    colorlinks=true,
    linkcolor=blue,
    filecolor=red,      
    urlcolor=magenta,
    breaklinks=true,            
}
\usepackage{breakurl}           
\usepackage{pifont}
\usepackage{array}
\usepackage{url}
\usepackage{dblfloatfix}
\usepackage{booktabs}
\usepackage{amsmath}
\usepackage{tabularx}
\usepackage{hyperref}
\usepackage{listings}
\usepackage{colortbl,array,xcolor}

\usepackage{graphics}
\usepackage{epsfig}
\usepackage{multirow}
\usepackage{url}
\usepackage{tikz}
\usepackage{enumitem}
\usepackage{multicol}

\usepackage{verbatim}
\usepackage{float}
\usepackage{subcaption}

\usepackage{algorithm}
\usepackage{algpseudocode}
\usepackage{graphicx}
\usepackage{calc}

\usepackage{marginnote}
\usepackage{xspace}

\definecolor{dark_green}{rgb}{0.0, 0.5, 0.0}
\definecolor{maroon}{cmyk}{0, 0.87, 0.68, 0.32}
\definecolor{halfgray}{gray}{0.55}
\definecolor{ipython_frame}{RGB}{207, 207, 207}
\definecolor{ipython_bg}{RGB}{247, 247, 247}
\definecolor{ipython_red}{RGB}{186, 33, 33}
\definecolor{ipython_green}{RGB}{0, 128, 0}
\definecolor{ipython_blue}{RGB}{64, 128, 128}
\definecolor{ipython_purple}{RGB}{170, 34, 255}
\usepackage{xcolor,colortbl}
\usepackage{minted}

\newcommand\blue[1]{\textcolor{blue}{#1}}

\newcommand{\custo}[1]{\textsc{\normalsize #1}}
\newcommand{\sprint}{\custo{SPRINT}\xspace}

\newcommand{\squishlist}{
 \begin{list}{$\bullet$}
  { \setlength{\itemsep}{0pt}
     \setlength{\parsep}{1pt}
     \setlength{\topsep}{1pt}
     \setlength{\partopsep}{0pt}
     \setlength{\leftmargin}{1.5em}
     \setlength{\labelwidth}{1em}
     \setlength{\labelsep}{0.5em} } }

\newcommand{\squishend}{
  \end{list}  }

\AtBeginDocument{%
  \providecommand\BibTeX{{%
    \normalfont B\kern-0.5em{\scshape i\kern-0.25em b}\kern-0.8em\TeX}}}

\copyrightyear{2023}
\acmYear{2023}
\setcopyright{acmlicensed}\acmConference[SIGIR '23]{Proceedings of the 46th International ACM SIGIR Conference on Research and Development in Information Retrieval}{July 23--27, 2023}{Taipei, Taiwan}
\acmBooktitle{Proceedings of the 46th International ACM SIGIR Conference on Research and Development in Information Retrieval (SIGIR '23), July 23--27, 2023, Taipei, Taiwan}
\acmPrice{15.00}
\acmDOI{10.1145/3539618.3591902}
\acmISBN{978-1-4503-9408-6/23/07}

\begin{document}
\sloppy

\title[SPRINT: A Unified Toolkit for Evaluating and Demystifying Zero-shot Neural Sparse Retrieval]{SPRINT: A Unified Toolkit for Evaluating and Demystifying Zero-shot Neural Sparse Retrieval}

\author{Nandan Thakur}
\affiliation{University of Waterloo \country{Waterloo, Canada}}

\author{Kexin Wang}
\affiliation{UKP Lab, TU Darmstadt \country{Darmstadt, Germany}}

\author{Iryna Gurevych}
\affiliation{UKP Lab, TU Darmstadt \country{Darmstadt, Germany}}

\author{Jimmy Lin}
\affiliation{University of Waterloo \country{Waterloo, Canada}}
\settopmatter{authorsperrow=4}

\renewcommand{\shortauthors}{Nandan Thakur, Kexin Wang, Iryna Gurevych, \& Jimmy Lin}

\begin{abstract}
Traditionally, sparse retrieval systems relied on lexical representations to retrieve documents, such as BM25, dominated information retrieval tasks. With the onset of pre-trained transformer models such as BERT, neural sparse retrieval has led to a new paradigm within retrieval. Despite the success, there has been limited software supporting different sparse retrievers running in a unified, common environment. This hinders practitioners from fairly comparing different sparse models and obtaining realistic evaluation results. Another missing piece is,
that a majority of prior work evaluates sparse retrieval models on in-domain retrieval, i.e. on a single dataset: MS MARCO. However, a key requirement in practical retrieval systems requires models that can generalize well to unseen out-of-domain, i.e. zero-shot retrieval tasks. In this work, we provide \sprint, a unified python toolkit based on Pyserini and Lucene, supporting a common interface for evaluating neural sparse retrieval. The toolkit currently includes five built-in models: uniCOIL, DeepImpact, SPARTA, TILDEv2 and SPLADEv2. Users can also easily add customized models by defining their term weighting method. Using our toolkit, we establish strong and reproducible zero-shot sparse retrieval baselines across the well-acknowledged benchmark, BEIR. Our results demonstrate that SPLADEv2 achieves the best average score of 0.470 nDCG@10 on BEIR amongst all neural sparse retrievers. 
In this work, we further uncover the reasons behind its performance gain. We show that SPLADEv2 produces sparse representations with a majority of tokens outside of the original query and document which is often crucial for its performance gains, i.e. a limitation among its other sparse counterparts. We provide our \sprint toolkit, models, and data used in our experiments publicly here: \blue{\href{https://github.com/thakur-nandan/sprint}{https://github.com/thakur-nandan/sprint}}.

\begin{table}[t]
    \vspace{7mm}
    \resizebox{0.45\textwidth}{!}{\begin{tabular}{l|c|c|c|c}
        \toprule
        \textbf{Retrieval} & \textbf{Lexical} & \textbf{Inverted Index} & \textbf{CPU for} & \textbf{Interpret-} \\ 
         \textbf{System} & \textbf{Matching} & \textbf{for Inference} & \textbf{Inference} & \textbf{ability} \\ \midrule
        Sparse & \textcolor{dark_green}{$\boldsymbol{\checkmark}$} & \textcolor{dark_green}{$\boldsymbol{\checkmark}$} & \textcolor{dark_green}{$\boldsymbol{\checkmark}$} & \textcolor{dark_green}{$\boldsymbol{\checkmark}$} \\
        Dense & \textcolor{red}{$\boldsymbol{\times}$} & \textcolor{red}{$\boldsymbol{\times}$} & \textcolor{red}{$\boldsymbol{\times}$} & \textcolor{red}{$\boldsymbol{\times}$} \\
        \bottomrule
    \end{tabular}}
        \caption{An overview of advantages of sparse retrieval systems inherited from Bag-of-Words (BoW) lexical models over its dense counterparts in information retrieval.}
    \label{tab:sparse-vs-dense}
    \vspace{-8mm}
\end{table}

\end{abstract}





\maketitle

\section{Introduction}
\label{sec:intro}

\begin{table*}[t]
    \small
    \resizebox{0.9\textwidth}{!}{\begin{tabular}{l!{\color{lightgray}\vrule} c!{\color{lightgray}\vrule} c!{\color{lightgray}\vrule} c!{\color{lightgray}\vrule} c! {\color{lightgray}\vrule} c!{\color{lightgray}\vrule} c}
        \toprule
        \textbf{Design Architecture} &
        \textbf{DeepImpact} &
        \textbf{TILDEv2} &
        \textbf{uniCOIL} &
        \textbf{SPARTA} &
        \textbf{SPLADEv2} &
        \textbf{BT-SPLADE-L} \\ \midrule \arrayrulecolor{lightgray}
        \multirow{2}{*}{Representation Building} & \multicolumn{1}{p{1.3cm}|}{MLP $\rightarrow scalar$} & \multicolumn{1}{p{1.3cm}|}{MLP $\rightarrow scalar$} & \multicolumn{1}{p{1.3cm}|}{MLP $\rightarrow scalar$} & \multicolumn{1}{p{1.3cm}|}{$|V_{\text{BERT}}|$ + Pooling} & \multicolumn{1}{p{1.3cm}|}{$|V_{\text{BERT}}|$ + Pooling} & \multicolumn{1}{p{1.3cm}}{$|V_{\text{DistilBERT}}|$ + Pooling} \\ \midrule
        \multicolumn{1}{p{3cm}|}{Document Expansion} & DocT5query$^\dagger$ & TILDE$^\dagger$ & DocT5query$^\dagger$ & SPARTA & SPLADEv2 & BT-SPLADE-L \\ \midrule
        \multicolumn{1}{p{3cm}|}{Hard Negatives} & BM25 & BM25 & BM25 & Random & SPLADE & SPLADE \\ \midrule 
        \multicolumn{1}{p{3cm}|}{Knowledge Distillation} & \textcolor{red}{$\boldsymbol{\times}$} & \textcolor{red}{$\boldsymbol{\times}$} & \textcolor{red}{$\boldsymbol{\times}$} & \textcolor{red}{$\boldsymbol{\times}$} & \textcolor{dark_green}{$\boldsymbol{\checkmark}$} & \textcolor{dark_green}{$\boldsymbol{\checkmark}$} \\ \midrule
        \multicolumn{1}{p{3cm}|}{Query Term Weighting} & \textcolor{red}{$\boldsymbol{\times}$} & \textcolor{red}{$\boldsymbol{\times}$} & \textcolor{dark_green}{$\boldsymbol{\checkmark}$} & \textcolor{red}{$\boldsymbol{\times}$} & \textcolor{dark_green}{$\boldsymbol{\checkmark}$} & \textcolor{dark_green}{$\boldsymbol{\checkmark}$} \\ \midrule
        \multicolumn{1}{p{3cm}|}{Regularization} & \textcolor{red}{$\boldsymbol{\times}$} & \textcolor{red}{$\boldsymbol{\times}$} & \textcolor{red}{$\boldsymbol{\times}$} & \textcolor{red}{$\boldsymbol{\times}$} & \textcolor{dark_green}{$\boldsymbol{\checkmark}$} & \textcolor{dark_green}{$\boldsymbol{\checkmark}$} \\ \midrule
        \multicolumn{1}{p{3cm}|}{Stop Words Exclusion} & \textcolor{dark_green}{$\boldsymbol{\checkmark}$} & \textcolor{dark_green}{$\boldsymbol{\checkmark}$} & \textcolor{red}{$\boldsymbol{\times}$} & \textcolor{red}{$\boldsymbol{\times}$} & \textcolor{red}{$\boldsymbol{\times}$} & \textcolor{red}{$\boldsymbol{\times}$} \\
        \arrayrulecolor{black}
        \bottomrule
    \end{tabular}}
        \vspace{1mm}
        \caption{A comparison of different neural sparse retrieval design architectures evaluated in our work. $|V_{\text{BERT}}|$ denotes the BERT vocabulary, similarly $|V_{\text{DistilBERT}}|$ denotes the DistilBERT vocabulary. ($\dagger$) denotes ``external'' document expansion techniques namely DocT5query or TILDE.}
    \vspace{-7mm}
    \label{tab:comparison}
\end{table*}

Information Retrieval is a core task for many web search and NLP applications. 
Traditionally retrieval systems involved sparse bag-of-words (BoW) models such as BM25 \cite{bm25} or TF-IDF relied upon the exact \textit{lexical} match and frequency-based term importance matching between the query and every document in a collection. 
However, these traditional lexical-based retrieval systems face challenges in \textit{vocabulary-mismatch} \cite{berger2000bridging} with searching semantically relevant documents. 
Recently, deep learning and in particular pre-trained Transformer models like BERT \cite{devlin-etal-2019-bert} have become popular in information retrieval \cite{lin2020pretrained}. Dense retrieval \cite{karpukhin-etal-2020-dense, lee-etal-2019-latent} method involves retrieving documents using semantic embedding similarities. Dense retrieval has been successfully applied across many retrieval-based tasks such as web search, question-answering \cite{karpukhin-etal-2020-dense, xiong2020approximate, Hofstaetter2021_tasb_dense_retrieval, qu-etal-2021-rocketqa, liang2020embeddingbased, ma2021zeroshot}. 
However, dense retrieval systems are inefficient. For large-sized corpora, dense models produce bulky indexes and require approximate nearest neighbor (ANN) solutions such as faiss \cite{johnson2019billion} for quick inference at scale. 
They are also computationally expensive to serve practically, as the index needs to be loaded completely within the RAM amounting to high hardware costs.

On the other hand, there has been a growing interest in learning \textbf{sparse} representations for both queries and documents in information retrieval \cite{10.1145/3269206.3271800}. 
Neural sparse retrieval systems either assign scalar weights to word pieces or pool token weights on BERT's \cite{devlin-etal-2019-bert} vocabulary space.
Similarity scores between query and document are computed either as exact token overlap or using max-pooling.
Many recent works consider the sparse retrieval setup suitable for retrieval-based tasks \cite{sparterm-2020, splade-v1, gao-etal-2021-coil, mallia2021learning, nogueira2019doc2query, zhao-etal-2021-sparta, formal2021splade, tilde-v1, zhuang2021fast, lin2021brief, slim}.
Sparse retrievers have several noted advantages over their dense counterparts as summarized in \autoref{tab:sparse-vs-dense}.
Overall, sparse retrievers have two major benefits: first, they are cost-effective requiring only CPU for inference.
Second, lexical matching provides a boost in their effectiveness in contrast to dense retrievers.

Despite previous work, there have been limited software tools that provide an easy out-of-box evaluation of different neural sparse retrievers. 
A unified toolkit would enable a fair comparison and better adoption of different sparse retrieval systems by practitioners. 
To solve this, in our work, we provide \textbf{SPRINT}\footnote{SPRINT Toolkit: \blue{\href{https://github.com/thakur-nandan/sprint}{https://github.com/thakur-nandan/sprint}}} (\textit{SParse RetrIeval Neural Toolkit}), a Python toolkit unifying evaluation of different neural sparse retrievers. 
The toolkit provides flexibility for the user to explore five popular built-in models: uniCOIL \cite{lin2021brief}, DeepImpact \cite{mallia2021learning}, SPARTA \cite{zhao-etal-2021-sparta}, TILDEv2 \cite{zhuang2021fast}, and SPLADEv2 \cite{formal2021splade}. 
The toolkit also lets you add customized sparse models by defining the term weights of input queries and documents. 
Unlike other retrieval toolkits, \sprint is one of the first to focus on sparse neural search. It is an easy-to-use, modular, and extensible framework written in Python. 
\sprint is well-integrated with Pyserini \cite{Lin_etal_SIGIR2021_Pyserini} and Apache Lucene, which is \emph{de facto} package for inverted index implementation across both academia and industry. 

Across several real-world retrieval use cases, the scarcity of training data is a prevalent issue. 
Due to which, retrieval systems are often applied in a zero-shot setup. 
Where, the retrieval model is trained on a publicly available massive training dataset such as MS MARCO \cite{nguyen2016ms} and evaluated across a target retrieval task with no available in-domain training data. 
Prior work on sparse retrieval except for the SPLADE family \cite{formal2021splade, bt-splade-l} has only been evaluated in-domain, i.e. on a single dataset. 
Therefore, the generalization capabilities of neural sparse models yet remain unexplored. 
In this work, using our \sprint toolkit, we conduct a range of ad-hoc retrieval benchmarking experiments and provide strong zero-shot sparse baselines on the well-acknowledged BEIR benchmark \cite{thakur2021beir} for 18 datasets spanning diverse domains and zero-shot tasks. 

Overall, this is one of the first works to evaluate a wide range of sparse retrievers and provide a fair comparison among each other on zero-shot retrieval. 
In addition to providing baseline scores using our toolkit, in the latter half of our work (Section \ref{sec:demystifying-sparse-retrieval}), we investigate several Research Questions (RQ) to provide insights into neural sparse retrievers which would benefit the retrieval community:

\newcommand{\RQone}{\begin{itemize}
    \item[\textbf{RQ1}] \emph{How effective are sparse retrievers across zero-shot IR tasks?}
\end{itemize}}
\RQone

\noindent
Sparse retrievers without external document expansion, with the exception of SPLADEv2 and BT-SPLADE-L, suffer from generalization and underperform BM25 on BEIR. 
On the other hand, docT5query-based document expansion is crucial for improving the generalization accuracy of both uniCOIL and TILDEv2 sparse models by +1.3 and +3.1 nDCG10 points on average across 18 BEIR datasets. 
SPLADEv2 achieves the highest accuracy of 0.470 nDCG@10 on BEIR outperforming BM25 on average by +4.1 points. 

\newcommand{\RQtwo}{\begin{itemize}
    \item[\textbf{RQ2}] \emph{Why does SPLADEv2 achieve the highest zero-shot accuracy gain on BEIR?}
\end{itemize}}
\RQtwo

\noindent
SPLADEv2 produces document representations with minimum sparsity, i.e. a majority of non-zero weighted tokens are not present within the original passage. 
These ``expanded'' tokens, we show are critical for SPLADEv2 to achieve its strong zero-shot generalization on BEIR. 
When evaluated without its document expansion tokens, SPLADEv2 underperforms the lower baseline uniCOIL (no expansion) model on 6 out of the 8 BEIR datasets. 

\newcommand{\RQthree}{\begin{itemize}
    \item[\textbf{RQ3}] \emph{Which document expansion technique: docT5query or TILDE?}
\end{itemize}}
\RQthree

\noindent
As two document expansion methods, although docT5query is $140 \times$ more computationally expensive than TILDE, for sparse models such as uniCOIL and TILDEv2, these two methods perform comparably.
However, docT5query is found to repeat keywords in its generated queries from passages. This assists as a strong signal for BM25, where docT5query overall outperforms TILDE.

\newcommand{\RQfour}{\begin{itemize}
    \item[\textbf{RQ4}] \emph{Which model provides the best practical performance (efficiency and effectiveness) on BEIR?}
\end{itemize}}
\RQfour

\noindent
Despite all neural models being much slower than BM25 (3.5$\times$ $\sim$46$\times$ slower), BT-SPLADE-L significantly outperforms BM25 on effectiveness (+6.3 nDCG@10 points on HotpotQA) while remaining relatively efficient (5$\times$ slower), achieving the best trade-off on the pareto frontier. 
Query latency is found to scale linearly with query length for all the compared models.

\paragraph{Contributions}
This work focuses on sparse retrieval evaluation in the zero-shot retrieval setup. We provide a resource toolkit, novel sparse-retrieval baseline scores on BEIR, and an in-depth analysis of zero-shot sparse retrieval. Overall, our work has the following series of contributions:

\begin{itemize}[leftmargin=*]
\item We share the \sprint toolkit with the IR and NLP community, which extends the evaluation of several neural sparse retrievers on a common interface.
The toolkit would enable a fair comparison of different sparse retrievers and allow practitioners to search on their custom datasets using sparse retrieval.

\item We present strong sparse retrieval baselines for zero-shot retrieval on BEIR using \sprint. 
As a part of the resource, we also release all our experiment data including inverted indexes and expanded BEIR passages.

\item We conduct experiments to study reasons why certain sparse models generalize well across zero-shot tasks in BEIR.
We explore topics such as document expansion and sparsity in depth and attempt to identify reasons for sparse model effectiveness.
\end{itemize}

\begin{figure*}[t!]
    \centering
    \includegraphics[width=0.425\textwidth,clip, trim=0 9 5 5]{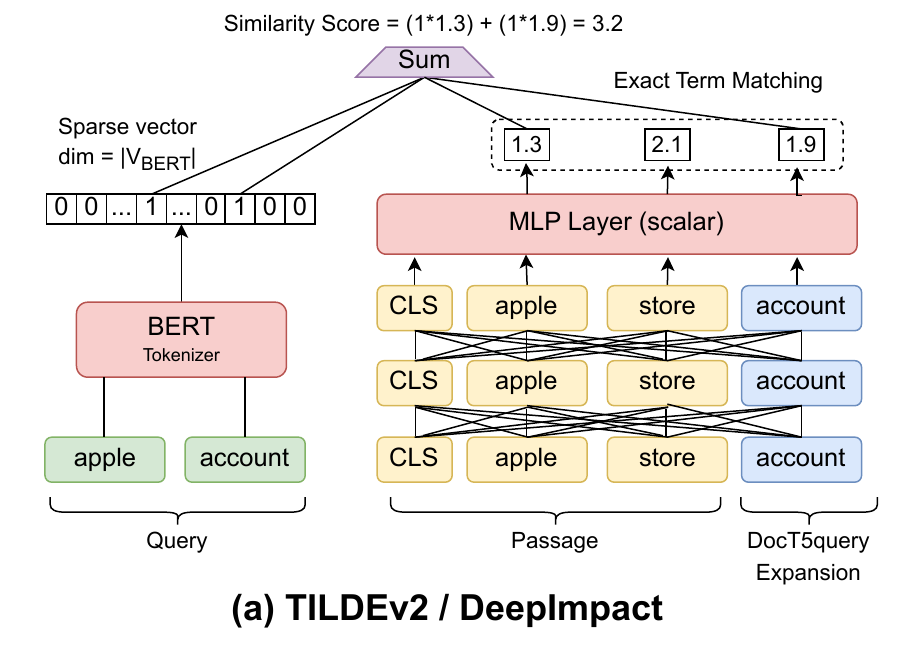}
    \hspace{0.5cm}
    \includegraphics[width=0.425\textwidth,clip, trim=0 11 0 5]{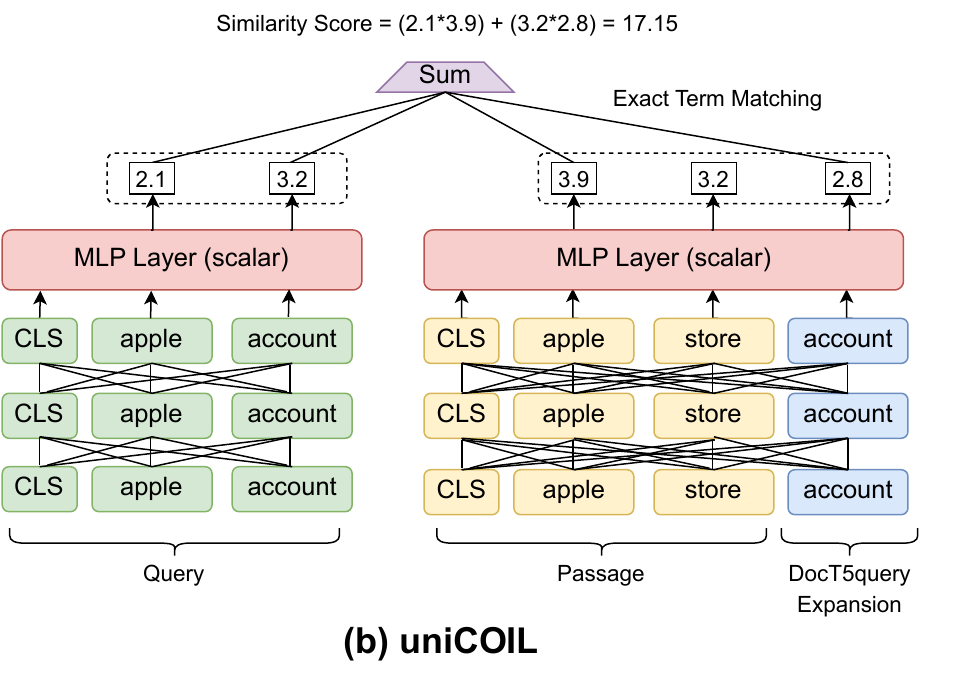} \\ \vspace{0.5cm}
    \includegraphics[width=0.425\textwidth,clip, trim=0 10 0 5]{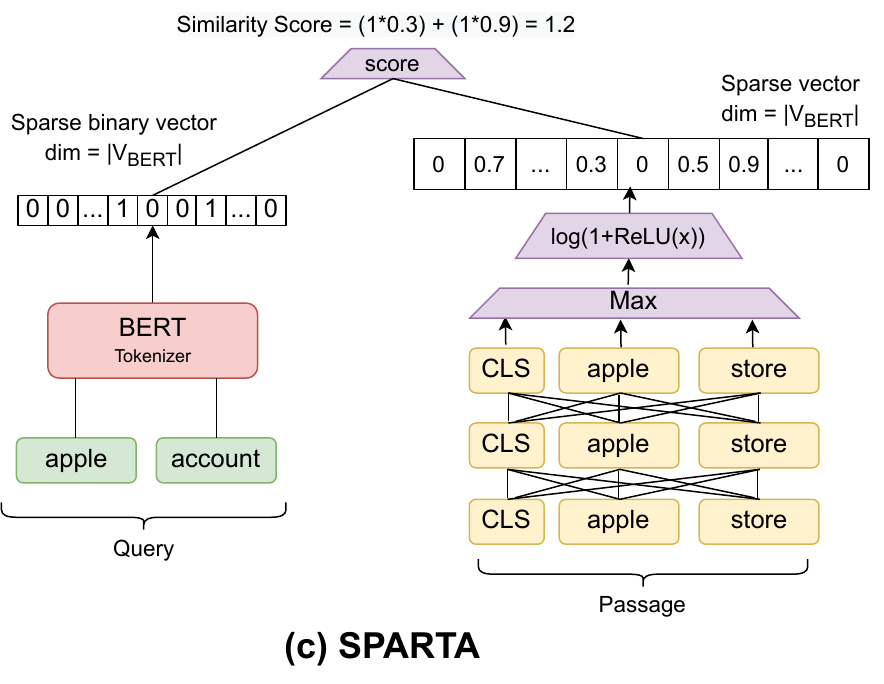}
    \hspace{0.5cm}
    \includegraphics[width=0.45\textwidth,clip, trim=0 10 0 5]{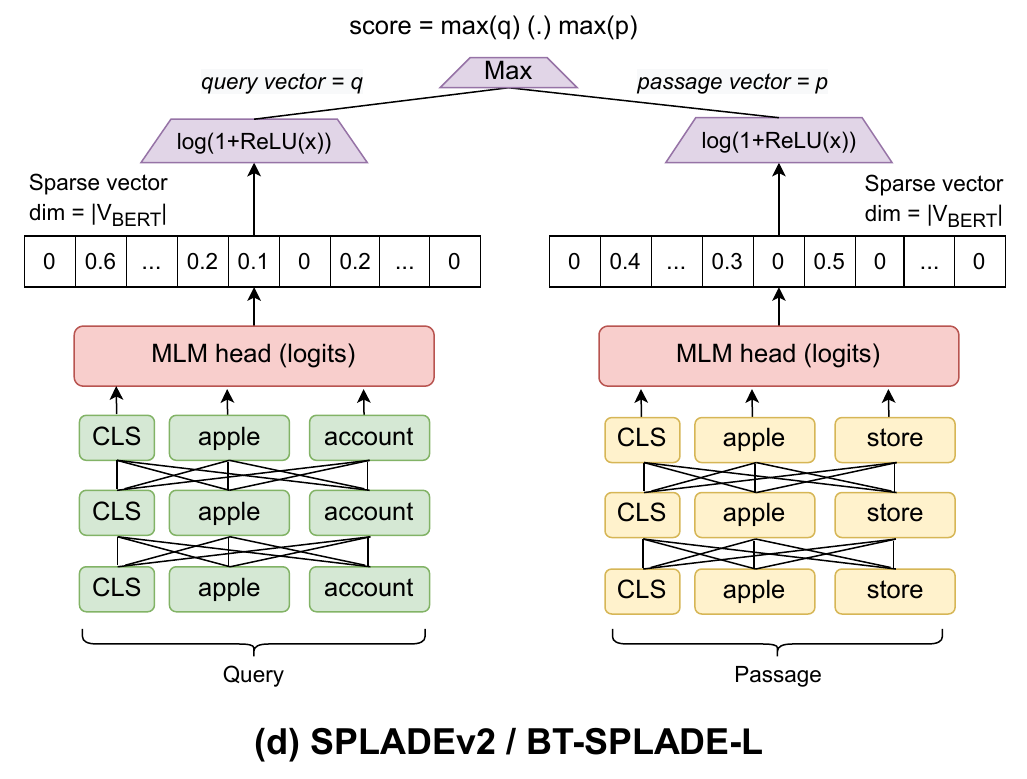}
  \caption{\textbf{A summary of neural sparse model architectures available in the \sprint toolkit. The expansion terms (blue color tokens) are generated using an external document expansion (docT5query) technique.}}
  \vspace{-0.25cm}
  \label{fig:sparse-architectures}
\end{figure*}

\section{Related Work}
\paragraph{Neural Sparse Retrieval}
Sparse models perform retrieval by term weighting based on BERT-based networks, which can go along with document and query expansion~\cite{sparterm-2020, splade-v1, gao-etal-2021-coil, mallia2021learning, nogueira2019doc2query, zhao-etal-2021-sparta, formal2021splade, tilde-v1, zhuang2021fast, lin2021brief, slim}. 
There are two categories of sparse models: The first category (including DeepImpact, TILDEv2 and uniCOIL) learns scalar weights for terms present within the passage \cite{lin2021brief, mallia2021learning, zhuang2021fast, tilde-v1}. 
These models by default do not perform either query or document expansion. On the other hand, the other category (including SPARTA, BT-SPLADE-L and SPLADEv2) pools token weights on BERT's vocabulary space \cite{formal2021splade, bt-splade-l, slim, zhao-etal-2021-sparta, sparterm-2020} and thus representing queries and documents with possibly different term tokens. 
As we have a wide variety of options available for sparse models, in our work, we focus on including the fundamental sparse architectures. 
Variants of these fundamental sparse architectures, such as for SPLADE \cite{densifying-sparse, sparsifying-sparse} or models with similar base architectures, such as SparTerm \cite{sparterm-2020}, TILDEv1 \cite{tilde-v1} or SPLADE \cite{splade-v1}, have been ommitted out for the sake of brevity.
A recent work on wacky weights \cite{wacky-weights} evaluated several sparse models together on the Pareto frontier on MSMARCO. We conduct a similar Pareto analysis for zero-shot retrieval for our work.
LSR \cite{lsr} is a recent concurrent work with ours. The authors provide a toolkit focused on sparse-retrieval model training and evaluate different training settings with in-domain datasets such as MS MARCO \cite{nguyen2016ms}. In contrast to prior work, we focus on evaluating a broad category of neural sparse retrievers on the zero-shot retrieval setting.

\paragraph{Zero-shot Sparse Baselines} SPLADEv2 originally provided zero-shot baselines for 13 datasets in BEIR \cite{formal2021splade}. 
They recently extended the baseline scores on all 18 datasets in BEIR \cite{bt-splade-l}. 
However, prior work in SPLADE misses a fair comparison against other recent sparse models in this zero-shot setting. 
In our work, we provide a comprehensive study to fairly evaluate all different sparse models in an identical evaluation setup using \sprint.

\paragraph{Related IR toolkits} Existing toolkits, with the exception of Pyserini, do not support search using neural sparse retrievers. These toolkits provide search systems consisting of lexical, dense retrieval, or reranker architectures. For example, Pyterrier \cite{pyterrier2020ictir} helps easy search across documents using dense retrievers or neural re-rankers. Caproleus \cite{10.1145/3340531.3412780} lets users create pipelines where they can combine lexical search and neural reranking. 
FlexNeuArt \cite{boytsov2020flexible} integrates with Lucene and NMSLIB \cite{DBLP:conf/sisap/BoytsovN13} and provides search primarily on dense and lexical-based retrievers. Lastly, companies providing neural-search solutions such as Naver have not publicly open-sourced their implementations.

\section{SPARSE Retrieval Models}
In this section, we provide an overview of different sparse model architectures. For an overview of the comparison of the design architectures, we would like to refer the reader to Table \ref{tab:comparison}.

\paragraph{DeepCT and DocT5query} DeepCT and DocT5query are the early sparse models that work with BERT. 
DeepCT \cite{10.1145/3397271.3401204} used a bert-base-uncased model finetuned on MS MARCO to learn the term weight frequencies (tf). 
The model generates a pseudo-document with keywords multiplied by the learned term-frequencies. 
On the other hand, DocT5query \cite{nogueira2019document} identified document expansion terms using a seq-to-seq encoder-decoder model such as T5 \cite{JMLR:v21:20-074} that generated top-$k$ possible queries for which the given document would be relevant. 
These generated queries would be appended to the original passage for the BM25-based lexical search.

\paragraph{TILDEv2 / DeepImpact} TILDEv2 \cite{zhuang2021fast} is an extension of TILDE \cite{tilde-v1}. 
The model learns to output a scalar weight for each token within the input passage. 
TILDEv2 architecture uses BERT \cite{devlin-etal-2019-bert} contextualized token embeddings for all passage tokens and feeds them into an additional projection (MLP) layer on top of BERT to downsize the 768-dimensional embeddings to a single dimension, i.e. scalar weight. 
The input query is only tokenized and a sparse binary vector is produced. 
To compute relevancy, the model performs exact-term matching between the intersection of tokens present in both the query and passage. 
The TILDEv2 model is trained on the MS MARCO dataset using InfoNCE~\cite{infoNCE} loss function. 
Each training batch consisted of hard negatives which were created by randomly sampling passages obtained from the top-1000 results obtained by BM25 for each input query in MS MARCO.
DeepImpact \cite{mallia2021learning} uses a similar architecture as TILDEv2 and was trained in an identical setup. The main difference lies in DeepImpact learns to effectively model word-level term weights, i.e. each dimension in the sparse representation is a word, while TILDEv2 (and also other sparse methods) rely on BERT token-level term weights. 
DeepImpact also uses BERT \cite{devlin-etal-2019-bert} contextualized token embeddings for all passage tokens. The model feeds the first token representation of the occurrence of each first word into a projection (MLP) layer to downsize the 768-dimensional embeddings to a single dimension. 
The final representation contains non-repeated words and the corresponding scalar weights for each word in the input passage. 
To compute relevancy, exact-term matching is performed.

\paragraph{uniCOIL}
uniCOIL \cite{lin2021brief} shares the identical architecture and training method as TILDEv2. 
The main difference between uniCOIL and existing TILDEv2 and DeepImpact is that uniCOIL learns to represent query term weights by sharing the encoder model between the query and document encodings.

\paragraph{SPARTA} 
SPARTA \cite{zhao-etal-2021-sparta} involves a different design architecture from previously discussed sparse models. 
The model architecture first gets BERT contextualized token embeddings $s_j (j=1, ..., l)$ for all passage tokens $x_j (j=1, ..., l)$. 
Then for each token $x_i (i=1,...,|V|)$ in the vocabulary $V$, it matches the passage token $s_j$ by maximizing the dot-product between the BERT input embedding $e_i$ of $x_i$ and the passage token embedding $s_j$, obtaining the matching score $y_i$:
$$
    y_i = \max_{j=1, ..., l}e_i^Ts_j.
$$
Next, these matching scores $y_i$'s will go through ReLU activation and log operation to get the final term weights $\phi(x_i) (i=1,..., |V|)$:
$$
    \phi_{\mathrm{SPARTA}}(x_i) = \log(\mathrm{ReLU}(y_i + b)+1),
$$
where $b$ is a learnable bias term. 
The passage representation can contain different terms from terms which appears in the passage. 
SPARTA represents the query representation as a sparse binary vector with terms present in the query with a weight equal to one. 
The original SPARTA model was trained on NQ \cite{47761}, however, we reuse the SPARTA implementation in \cite{thakur2021beir}, which was trained using the InfoNCE loss function on MS MARCO, with a negative passage set created by randomly sampling passages from the collection.

\paragraph{SPLADEv2 / BT-SPLADE-L}
SPLADEv2 \cite{formal2021splade} models the interaction between all BERT input embeddings and the BERT output token representations. The main difference is, SPLADEv2 also models query term weights with a shared encoder and changes the specific representation-building process into using an MLM layer followed by max-pooling. The model first gets BERT contextualized token embeddings $s_j (j=1, ..., l)$ for all passage tokens $x_j (j=1, ..., l)$. Then it maps these passage token embeddings into token-prediction logits $w_{i,j} (i=1, ..., |V|)$ via a MLM layer:
$$
  w_{i,j} = \mathrm{MLM}(s_j)[i],
$$
where $[i]$ represents taking the element at the $i$-th position from a vector. The final text representation is obtained via ReLU activation, log operation and max-pooling:
$$
    \phi_{\mathrm{SPLADEv2}}(x_i) = \max_{j=1, ..., l}\log(\mathrm{ReLU}(w_{i,j}+1))
$$

The SPLADEv2 model setup involved a knowledge distillation \cite{knowledge-distillation} setup with a cross-encoder as the teacher model \cite{hofstatter2021improving}. 
The model was trained MarginMSE loss \cite{hofstatter2021improving} and FLOPs regularization \cite{Paria2020Minimizing}. 
To mine hard negatives for MarginMSE training, it first trains a SPLADEv2 model with InfoNCE loss and FLOPs regularization using BM25-mined negatives and in-batch negatives. 
BT-SPLADE-L \cite{bt-splade-l} was a extension work in an attempt to make the existing SPLADEv2 model better in terms of efficiency. 
It shares the identical architecture and training method as SPLADEv2. 
The main difference between BT-SPLADE-L and existing SPLADEv2 is that the query and document encoder are separate in BT-SPLADE-L and the query encoder is much smaller than the document encoder, which helps in reducing the query latency significantly in BT-SPLADE-L.

\section{Sprint Toolkit Overview}

For a fair comparison of different sparse retrievers, models must be evaluated in an identical setup. Existing retrieval toolkits focus either on lexical, dense, or reranking-based retrieval setups. 
To support sparse (neural) search, we developed \sprint (\textbf{SP}arse \textbf{R}etr\textbf{I}eval \textbf{N}eural \textbf{T}oolkit)\footnote{Throughout the paper we have used the term: ``neural sparse retrieval'', However above for the acronym purposes, we refer SPRINT as ``sparse retrieval neural toolkit''.} is an easy-to-use Python toolkit, focused completely on neural sparse search. 
Using \sprint, practitioners can effectively search on their own datasets using different sparse retrievers. \sprint integrates effectively around Pyserini (Section \ref{sec:pyserini}), a Lucene-based index implementing an inverted index implementation for efficient sparse retrieval. Currently, the toolkit supports out-of-the-box evaluation of five sparse models: uniCOIL \cite{lin2021brief}, DeepImpact \cite{mallia2021learning}, SPARTA \cite{zhao-etal-2021-sparta}, TILDEv2 \cite{zhuang2021fast}, and SPLADEv2 \cite{formal2021splade}. The toolkit also integrates support from Hugging Face,\footnote{\url{https://huggingface.co/}} as model checkpoints can be directly loaded from the HF hub \cite{wolf2020huggingfaces} within the toolkit and evaluated across any publicly available dataset in the BEIR benchmark (Section \ref{sec:beir}). \sprint improves Pyserini for sparse retrieval in the following ways: (1) \sprint contains the complete end-to-end inference pipeline including all data preparation steps within the toolkit which in Pyserini is not supported. (2) A user can easily use \sprint to evaluate their own sparse model architecture, by providing the query and document encoder class to provide term weights for the input query and passage respectively.

\begin{figure}[t!]
    \centering
    \includegraphics[width=0.45\textwidth]{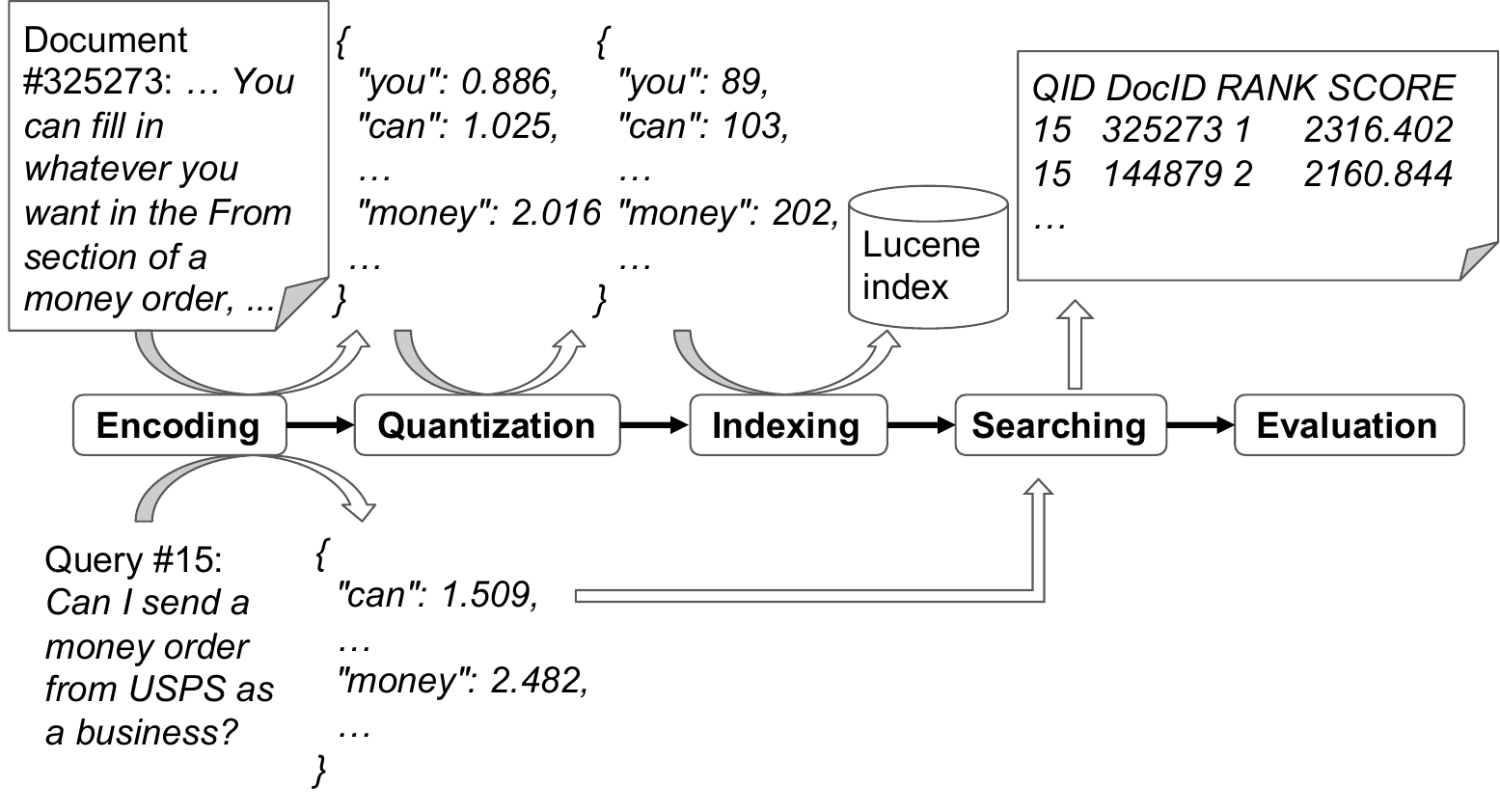}
  \vspace{-4mm}
  \caption{\textbf{Illustration of the five-step sequential inference pipeline in SPRINT. The example comes from the FiQA dataset \cite{10.1145/3184558.3192301} within the BEIR benchmark \cite{thakur2021beir}.}}
  \label{fig:ppl}
  \vspace{-4mm}
\end{figure}

\subsection{Inference Pipeline}
\sprint performs inference with a five-step sequential pipeline. As shown in~\autoref{fig:ppl}, we start with encoding documents into terms, i.e. token weights by multiprocessing on multiple GPUs. This helps in encoding large-sized corpora quickly, often containing millions of passages. 
Next, we quantize the float document term weights into integers, which are commonly adopted by the neural sparse retrievers. 
This reduces the model precision but is crucial for reducing the memory costs of the inverted index.
After the quantization step, we index the term weights into a Lucene index (backed by Pyserini). 
After the indexing step, we search by encoding query tokens to retrieve relevant documents from the Lucene index (backed by Pyserini). 
Finally, we evaluate the search results against certain gold relevance judgements, e.g. \textit{qrels} present within BEIR.

\subsection{Example Usage}
To simplify the complexity of the above inference for the user, we provide an inference module named ~~\texttt{aio}, a short form for ``\textit{all-in-one}'' for starting the inference pipeline in \sprint. 
Below we provide a starter code example from \sprint to evaluate the uniCOIL (no expansion) model on the test split from the Scifact \cite{wadden-etal-2020-fact} dataset within the BEIR benchmark \cite{thakur2021beir}.

\begin{minted}[fontsize=\small]{python}
from sprint.inference import aio

if __name__ == '__main__':
    aio.run(
        encoder_name='unicoil',
        ckpt_name='castorini/unicoil-noexp-msmarco-passage',
        data_name='beir/scifact',
        gpus=[0, 1],
        output_dir='beir_scifact-unicoil_noexp',
        do_quantization=True,
        quantization_method='range-nbits',
        original_score_range=5,
        quantization_nbits=8,
        original_query_format='beir',
        topic_split='test'
    )
\end{minted}

\vspace{-3mm}
\subsection{Pyserini Integration}
\label{sec:pyserini}  
The established approach to evaluating dense retrievers at scale involves approximate nearest-neighbor search solutions such as faiss \cite{johnson2019billion}. 
Similarly, evaluating sparse retrievers at scale would require an efficient inverted index implementation. Here, Pyserini \cite{Lin_etal_SIGIR2021_Pyserini} which is based on Apache Lucene is becoming the de-facto standard. 
Pyserini allows easy evaluation and multi-thread encoding for indexing documents into a Lucene index. 
Pyserini only requires a vector of keywords present and their corresponding weights for indexing sparse tokens. 
Following previous work in \cite{wacky-weights}, we also investigated other inverted implementations such as PISA \cite{MSMS2019}. 
However, we found that Apache Lucene is better adapted within both research  and industry communities, while PISA is limited in research usage. 
In the future, we can look into integrating PISA and providing the user the option to choose their preferred inverted index implementation. 

\subsection{BEIR/Dataset Integration}
\label{sec:beir}
The BEIR benchmark provides a diverse set of retrieval tasks and datasets for evaluation. 
All the BEIR datasets follow an identical format, consisting a \textit{passage-level corpus}, \textit{test queries}, and \textit{qrels, i.e. a relevance judgements} file. 
\sprint currently extends evaluation on all BEIR datasets. In order to stay relevant in the future, the toolkit can be easily extended to newer novel datasets. 
A new dataset is required to be converted into the BEIR format.\footnote{{How to convert a custom dataset into BEIR format: \url{https://github.com/beir-cellar/beir/wiki/Load-your-custom-dataset}}} 
Once converted, the new dataset can be directly used within the SPRINT toolkit.

\section{Experimental Design}

\subsection{Datasets}
First, we conduct experiments on in-domain retrieval datasets to measure the reproducibility of our \sprint toolkit. 
The differences in scores would help identify sources of error during reproduction. 
From prior work, all sparse retrievers except SPARTA, have been fine-tuned on the MS MARCO dataset. 
The majority of the models were evaluated on MSMARCO-DEV \cite{nguyen2016ms} and the TREC-DL 2019 \cite{trec-dl-2019} track. 
In our work, we took the original model checkpoints, if available. Following standard practice, we report MRR@10 for MSMARCO-DEV and nDCG@10 for TREC-DL'19. 

Next, we use the BEIR benchmark, introduced by \citet{thakur2021beir}, which contains 18 retrieval datasets, corresponding to nine tasks, such as fact verification or question answering, and covering different domains, suchas news articles, Wikipedia, or scientific publications. 
Most datasets from BEIR do not contain a training set and focus on zero-shot retrieval. 
Following standard practice, we report nDCG@10 on the BEIR benchmark. 

\begin{table}[t]
    \centering
    \setlength\tabcolsep{3pt}
    \begin{tabular}{l!{\color{lightgray}\vrule}r!{\color{lightgray}\vrule}r!{\color{lightgray}\vrule}rr!{\color{lightgray}\vrule}r!{\color{lightgray}\vrule}r!{\color{lightgray}\vrule}r}
       
       \toprule

       \multirow{2}{*}{\textbf{Model}} &
       \multicolumn{3}{c}{\textbf{DEV (MRR@10)}} & &
       \multicolumn{3}{c}{\textbf{DL'19 (nDCG@10)}} \\
       \cmidrule(lr){2-4}
       \cmidrule(lr){6-8}
       & Orig. & {\small SPRINT} & Error & & Orig. & {\small SPRINT} & Error \\

       \midrule
       \arrayrulecolor{lightgray}
        BM25    & 0.187 & 0.187 & +0.0\% & & 0.519 & 0.519 & +0.0\%\\
        uniCOIL & 0.351 & 0.348 & -0.8\% & & 0.693 & 0.700 & +1.0\% \\
        SPLADEv2 & 0.368 & 0.368 &  +0.0\% & & 0.729 & 0.728 & -0.1\% \\
        BT-SPLADE-L & 0.380 & 0.378 & -0.5\% &  & 0.703 & 0.703 & +0.0\% \\
        DeepImpact & 0.326 & 0.327 & +0.3\% & & 0.695 & 0.709 & +2.0\% \\
        TILDEv2 & 0.333 & 0.331 & -0.6\% & &  0.676 & 0.657 & -2.8\% \\
        \arrayrulecolor{black}
       \bottomrule
    \end{tabular}
    \caption{Reproduction study of sparse neural models on in-domain datasets: MS MARCO DEV \cite{nguyen2016ms} and TREC DL'19 \cite{trec-dl-2019}.}
    \vspace{-4mm}
    \label{tab:reproduction}
\end{table}

\vspace{-2mm}
\subsection{Baselines}
In this subsection, we briefly introduce the different categories of baselines. 
We define ``Lexical baselines'' as models which rely upon BM25 for retrieval. 
Whereas, ``Sparse baselines'' rely on a Transformer-based model design producing sparse representations.

\begin{table*}[t]
    \centering
    \small    
    \setlength\tabcolsep{3pt}
    \begin{tabular}{l|ccccccc|ccccc}
       
       \toprule

       \textbf{Dataset ($\downarrow$)} &
       \multicolumn{7}{c!{\color{lightgray}\vrule}}{\underline{\textbf{Without Expansion}}}&
       \multicolumn{4}{c}{\underline{\textbf{With docT5query ($q=20$) Expansion}}}\\
       \multirow{2}{*}{Model ($\rightarrow$)} & \multirow{2}{*}{BM25$^*$} & \multirow{2}{*}{DeepCT$^*$} & \multirow{2}{*}{SPARTA$^*$} & \multirow{2}{*}{uniCOIL}  & BM25 (1k) & SPLADEv2 & BT-SPLADE & \multirow{2}{*}{DocT5} & \multirow{2}{*}{DeepImpact} & \multirow{2}{*}{uniCOIL} & DocT5 (100) \\
       && & &  & + TILDEv2$^\dagger$ & (distil) & (L/Large) & & & & + TILDEv2$^\dagger$ \\

        \midrule
       \arrayrulecolor{lightgray}
       TREC-COVID     & 0.656 &  0.406  & 0.538  & 0.640 & 0.621 & 0.710 & 0.661 & 0.682 & 0.674 & 0.710 & 0.716 \\ 
       BioASQ         & 0.465 &  0.407  & 0.351  & 0.477 & 0.469 & 0.508 & 0.471 & 0.487 & 0.467 & 0.488 & 0.471 \\ 
       NFCorpus       & 0.325 &  0.283  & 0.301  & 0.333 & 0.314 & 0.334 & 0.331 & 0.330 & 0.312 & 0.336 & 0.318 \\ \midrule
       NQ             & 0.329 &  0.188  & 0.398  & 0.425 & 0.396 & 0.521 & 0.515 & 0.377 & 0.433 & 0.467 & 0.410 \\ 
       HotpotQA       & 0.603 &  0.503  & 0.492  & 0.667 & 0.663 & 0.684 & 0.666 & 0.624 & 0.650 & 0.661 & 0.650 \\ 
       FiQA-2018      & 0.236 &  0.191  & 0.198  & 0.289 & 0.255 & 0.336 & 0.318 & 0.277 & 0.266 & 0.298 & 0.266 \\ \midrule
       Signal-1M (RT) & 0.330 &  0.269  & 0.252  & 0.275 & 0.273 & 0.266 & 0.283 & 0.277 & 0.274 & 0.280 & 0.284 \\ \midrule
       TREC-NEWS      & 0.398 &  0.220  & 0.258  & 0.374 & 0.304 & 0.392 & 0.387 & 0.395 & 0.303 & 0.358 & 0.341 \\ 
       Robust04       & 0.407 &  0.287  & 0.276  & 0.403 & 0.357 & 0.468 & 0.407 & 0.427 & 0.347 & 0.422 & 0.382 \\ \midrule
       ArguAna        & 0.414 &  0.309  & 0.279  & 0.387 & 0.351 & 0.478 & 0.474 & 0.463 & 0.374 & 0.373 & 0.476 \\
       Touch\'e-2020  & 0.367 &  0.156  & 0.175  & 0.298 & 0.296 & 0.272 & 0.270 & 0.234 & 0.311 & 0.297 & 0.269 \\ \midrule
       CQADupstack    & 0.299 &  0.268  & 0.257  & 0.301 & 0.291 & 0.350 & 0.330 & 0.281 & 0.292 & 0.322 & 0.302 \\ 
       Quora          & 0.789 &  0.691  & 0.630  & 0.663 & 0.510 & 0.838 & 0.723 & 0.776 & 0.677 & 0.754 & 0.710 \\ \midrule
       DBPedia        & 0.313 &  0.177  & 0.314  & 0.338 & 0.313 & 0.435 & 0.405 & 0.333 & 0.344 & 0.361 & 0.321 \\ \midrule
       SCIDOCS        & 0.158 &  0.124  & 0.126  & 0.144 & 0.141 & 0.158 & 0.153 & 0.163 & 0.146 & 0.145 & 0.146 \\ \midrule
       FEVER          & 0.753 &  0.353  & 0.596  & 0.812 & 0.734 & 0.786 & 0.749 & 0.790 & 0.771 & 0.799 & 0.756 \\ 
       Climate-FEVER  & 0.213 &  0.066  & 0.082  & 0.182 & 0.159 & 0.235 & 0.189 & 0.212 & 0.190 & 0.191 & 0.186 \\ 
       SciFact        & 0.665 &  0.630  & 0.598  & 0.686 & 0.650 & 0.693 & 0.674 & 0.670 & 0.633 & 0.679 & 0.647 \\
       \arrayrulecolor{black}
       \midrule
       Average & 0.429 & 0.307 & 0.340 & 0.428 & 0.394 & 0.470 & 0.445 & 0.433 & 0.415 & 0.441 & 0.425 \\ 
        \arrayrulecolor{black}
       \bottomrule
    \end{tabular}
    \caption{Zero-shot baselines of sparse neural retriever and lexical (BM25) models on the BEIR Benchmark using \sprint. (*) denotes the scores of BM25, DeepCT and SPARTA reproduced and verified from the original BEIR paper. ($\dagger$) indicates sparse models which are used at the reranking stage. For both uniCOIL and TILDEv2, docT5query (q=20) expansion and without expansion are differently trained models.}
    \label{tab:zero-shot-baselines}
    \vspace{-0.5cm}
\end{table*}

\paragraph{Lexical Baselines} BM25 \cite{bm25} is a popular bag-of-words (BoW) scoring function based on the word, i.e. lexical overlap between query and document. 
We implement BM25 using Pyserini with default parameters ($k_1=0.9$ and $b=0.4$). 
Following previous work in \cite{thakur2021beir}, we index the title (if available) and passage both with equal importance. 
DocT5query \cite{nogueira2019doc2query} is a T5-base \cite{JMLR:v21:20-074} model\footnote{\href{https://huggingface.co/BeIR/query-gen-msmarco-t5-base-v1}{https://huggingface.co/BeIR/query-gen-msmarco-t5-base-v1}} fine-tuned on MS MARCO \cite{nguyen2016ms} used for generating $k$ synthetic queries from a passage input. 
In this work, to save on computational costs, we generate 20 (instead of 40) queries for every dataset in BEIR. 
DeepCT \cite{10.1145/3397271.3401204} is a \texttt{bert-base-uncased} model fine-tuned on MS MARCO \cite{nguyen2016ms} to learn the term weight frequencies (tf). 
We reuse the setup of DeepCT used in previous work \cite{10.1145/3397271.3401204, thakur2021beir}.

\paragraph{Sparse Baselines} SPARTA \cite{zhao-etal-2021-sparta} computes the contextualized results in a 30k dimensional sparse vector. 
We reuse the setup of SPARTA in BEIR \cite{thakur2021beir}, which uses a \texttt{distilbert-base-uncased} \cite{sanh2020distilbert} model\footnote{\href{https://huggingface.co/BeIR/sparta-msmarco-distilbert-base-v1}{https://huggingface.co/BeIR/sparta-msmarco-distilbert-base-v1}} fine-tuned on MS MARCO. 
Next, uniCOIL \cite{lin2021brief} is a \texttt{bert-base-uncased} model which learns to output scalar weights for terms within the input passage. 
We evaluate uniCOIL in two zero-shot settings: (1) \texttt{unicoil-noexp}:\footnote{\href{https://huggingface.co/castorini/unicoil-noexp-msmarco-passage}{https://huggingface.co/castorini/unicoil-noexp-msmarco-passage}} fine-tuned on MS MARCO without document expansion and (2) \texttt{unicoil}:\footnote{\href{https://huggingface.co/castorini/unicoil-msmarco-passage}{https://huggingface.co/castorini/unicoil-msmarco-passage}} fine-tuned on MS MARCO using docT5query \cite{nogueira2019doc2query} expansion with 40 synthetic queries. 
DeepImpact \cite{mallia2021learning} is a \texttt{bert-base-uncased} model fine-tuned in a setup similar to uniCOIL with a difference of effectively learning word-level term weights instead of token-level.
We evaluate the original model checkpoint\footnote{\href{https://public.ukp.informatik.tu-darmstadt.de/kwang/sparse-retrieval/checkpoints/deepimpact-bert-base.zip}{https://public.ukp.informatik.tu-darmstadt.de/kwang/sparse-retrieval/checkpoints/deepimpact-bert-base.zip}} which was fine-tuned on MS MARCO with its own document expansion technique. 
TILDEv2 \cite{zhuang2021fast} is a \texttt{bert-base-uncased} model trained in a setup similar to uniCOIL. 
We re-implement the approach in \cite{zhuang2021fast} and use a hashtable to implement the index for ``reranking'' the search results from BM25 search. 
Following prior work in \cite{zhuang2021fast}, we evaluate TILDEv2 in two zero-shot settings: (1) \texttt{tildev2-noexp}:\footnote{\href{https://huggingface.co/ielab/TILDEv2-noExp}{https://huggingface.co/ielab/TILDEv2-noExp}} fine-tuned on MS MARCO without document expansion and reranks the top-1000 BM25 results and (2) \texttt{tildev2}:\footnote{\href{https://huggingface.co/ielab/TILDEv2-docTquery-exp}{https://huggingface.co/ielab/TILDEv2-docTquery-exp}} fine-tuned on MS MARCO with docT5query expansion, where the model reranks the top 100 docT5query results. 
SPLADEv2 \cite{formal2021splade} is a \texttt{distilbert-base-uncased} model using max pooling fine-tuned on MS MARCO (with hard negatives) in a knowledge distillation setup.
For our experiments, we use the \texttt{distilsplade-max} model\footnote{\href{https://download-de.europe.naverlabs.com/Splade_Release_Jan22/distilsplade_max.tar.gz}{https://download-de.europe.naverlabs.com/Splade\_Release\_Jan22/distilsplade\_max.tar.gz}} provided by authors in \cite{formal2021splade}. 
BT-SPLADE-L \cite{bt-splade-l} is an extension of the SPLADEv2 model. The model was trained to improve the efficiency of SPLADEv2. BT-SPLADE-L has a separate query and document encoder, where the query encoder is a BERT-tiny \cite{jiao-etal-2020-tinybert} model\footnote{\href{https://huggingface.co/naver/efficient-splade-VI-BT-large-query}{https://huggingface.co/naver/efficient-splade-VI-BT-large-query}} and the document encoder is a \texttt{distilbert-base-uncased} \cite{sanh2020distilbert} model.\footnote{\href{https://huggingface.co/naver/efficient-splade-VI-BT-large-doc}{https://huggingface.co/naver/efficient-splade-VI-BT-large-doc}}

\vspace{-1mm}
\subsection{Evaluation Settings}
All our experiments have been conducted using a maximum of four NVIDIA RTX A6000 GPUs. In this work, we did not have to retrain any neural sparse retriever.

\paragraph{Quantization} In order to store weights efficiently in memory within an inverted index, a quantization step is performed to avoid storing floating-point scores for each token. 
The quantized scores belong in the range of $[1,2^b - 1]$, where $b$ denotes the bits required to store the value. 
This quantization technique is used by all sparse models except SPLADEv2 and BT-SPLADE-L, which use an simpler quantization technique by rounding off the score $\times 100$.

\paragraph{Document Expansion} Sparse models such as uniCOIL, TILDEv2 and DeepImpact rely on external document expansion techniques for better sparse representations. 
For a realistic evaluation, our evaluation setup in BEIR must be identical to the training setup. 
Hence, for the evaluation of models with expansion, in this work, we generated 20 queries using docT5query and use them as expanded passages in BEIR. 
To save others from recomputing, we provide our expanded passages publicly here for all BEIR datasets.\footnote{Under datasets in HF Hub: \href{https://huggingface.co/income}{https://huggingface.co/income}}

\section{Experimental Results}
In this section, we evaluate how retrieval models perform across in-domain and generalize on out-of-domain tasks in BEIR.

\paragraph{In-domain results} From \autoref{tab:reproduction}, we observe on average similar results between models evaluated using \sprint toolkit with their original implementations. 
We were able reliably reproduce results for all sparse retrievers within an error margin of less than 1\% MRR@10 for MS MARCO DEV. 
For TILDEv2, we observe the maximum difference, this is likely due to the authors' BM25 setting in \cite{zhuang2021fast} was different.\footnote{We opened up a GitHub issue here: \href{https://github.com/ielab/TILDE/issues/1}{https://github.com/ielab/TILDE/issues/1}} 
From the results, all sparse models likely outperform BM25 on in-domain retrieval and BT-SPLADE-L achieves the maximum score of 0.380 MRR@10 on MSMARCO DEV. 

\paragraph{BEIR results} From Table \ref{tab:zero-shot-baselines}, we first observe that without (external) document expansion, only SPLADEv2, and BT-SPLADE-L can significantly outperform the BM25 baseline on BEIR for out-of-domain datasets. 
This actually challenges the generalization ability of these neural sparse retrieval models, which needs to be well addressed in future model development. 
Next, we find that document expansion is crucial for neural sparse retrievers to improve their generalization performance on BEIR. 
Adding document expansion improves the performance of both uniCOIL and TILDEv2 by 3.0\% and 7.9\% on BEIR respectively. 
Finally, SPLADEv2 overall achieves 0.470 nDCG@10, which is the current state-of-the-art performance among sparse retrievers and its counterparts on BEIR, outperforming BM25 by an outstanding of +9.6\% nDCG@10.

\begin{figure*}[t!]
\centering
\begin{center}
    \includegraphics[trim=0 0 0 0,clip,width=\textwidth]{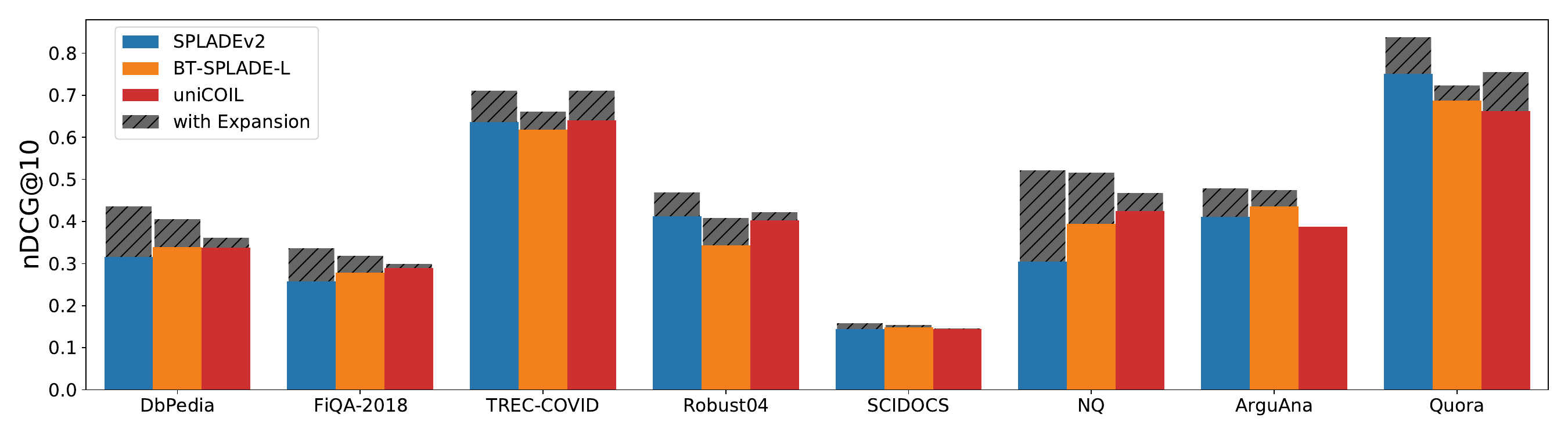}
    \vspace{-8mm}
    \captionof{figure}{nDCG10 scores of three neural sparse retrievers: SPLADEv2 (left, blue), BT-SPLADE-L (middle, orange), and uniCOIL (right, red) without (bold) and with document expansion (bold + hatch) on BEIR. Overall SPLADEv2 and BT-SPLADE-L either underperform or perform on par with uniCOIL, if document expansion tokens from each passage are removed.}
    \label{fig:splade-wo-expansion}
    \vspace{-0.25cm}
\end{center}
\end{figure*}

\begin{table}[t]
    \centering 
    \vspace{-0.2cm}
    \resizebox{\columnwidth}{!}{%
    \begin{tabular}{c|l|c!{\color{lightgray}\vrule}c!{\color{lightgray}\vrule}c!{\color{lightgray}\vrule}c}
       \toprule
       \textbf{Avg. Length} &
       \textbf{Model}   &
       \multicolumn{1}{c!{\color{lightgray}\vrule}}{\textbf{HotP.}}&
       \multicolumn{1}{c!{\color{lightgray}\vrule}}{\textbf{SciFact}}&
       \multicolumn{1}{c!{\color{lightgray}\vrule}}{\textbf{FiQA}} & \multicolumn{1}{c}{\textbf{NFC.}} \\

       \midrule
        \multirow{4}{*}{\textbf{Query}} & \textbf{SPLADEv2}   & 43.3 & 111.9 & 38.6 & 24.3 \\
                               & \textbf{BT-SPLADE-L} &13.1 & 15.7  & 9.5  & 4.6 \\ 
                               & \textbf{uniCOIL}   &19.1  & 20.6 & 13.9 & 5.9 \\
                               & \textbf{DeepImpact} &10.4  & 9.1  & 7.2 &  2.9 \\
       \midrule
       \multirow{4}{*}{\textbf{Passage}} & \textbf{SPLADEv2}    &197.8 &  389.1 & 318.6  & 379.2 \\
                                 & \textbf{BT-SPLADE-L} & 142.1 & 384.3 & 282.2 & 376.1 \\ 
                                 & \textbf{uniCOIL} &57.3 & 166.4 & 109.9 &  173.7 \\
                                 & \textbf{DeepImpact} &51.1 & 118.0 & 101.8 & 123.7 \\
       \bottomrule
    \end{tabular}}
    \caption{Sparsity analysis by computing the average non-zero token weights present in query and passage representations produced by neural sparse retrievers on four BEIR datasets: HotpotQA, SciFact, FiQA and NFCorpus.}
    \vspace{-0.75cm}
    \label{tab:sparsity-analysis}
\end{table}

\section{Demystifying Sparse Retrieval}
\label{sec:demystifying-sparse-retrieval}

\RQtwo
Previously in Table \ref{tab:zero-shot-baselines}, we observe that SPLADEv2 achieves the highest zero-shot nDCG@10 score in BEIR amongst its sparse model counterparts. 
In this section, we attempt to identify the source of its achieved performance gains. 

Recall that SPLADEv2 learns to expand documents and queries during training. 
This means that the final passage or query representation might contain different terms or words from what originally appears in the passage or query.
To get a better understanding, We conduct a sparsity analysis by computing the average query and passage token length of the representations produced by sparse models. 
Sparsity calculates of the number of zeros in the sparse representation. Models with higher sparsity, i.e. lower density are more efficient in comparison due to fewer tokens are required to store within the index. Also, query latency is lower due to faster token overlap calculation. 

From Table \ref{tab:sparsity-analysis}, we observe that for both query and document, SPLADEv2 performs the poorest, in terms of sparsity, whereas DeepImpact has the highest sparsity. 
For query representation, SPLADEv2 can produce up to $\sim$111.9 average non-zero tokens for a query representation in SciFact. 
For document expansion, SPLADEv2 and BT-SPLADE-L produce approximately 2.4$\times$ and 3.2$\times$ more tokens on average than uniCOIL (with expansion) and DeepImpact, respectively. 
The sparsity analysis experiment highlights the observation that a majority of tokens in SPLADEv2's sparse representation contain  ``expansion'' tokens that are not often present within the original passage or query. 

Naturally, the next question which arises is how much the expansion terms contribute towards SPLADEv2's strong zero-shot performance on BEIR. 
To answer this, we conduct an ablation study where we evaluate SPLADEv2 without its own document expansion, i.e. the model cannot use expansion tokens and is restricted to only use tokens present in the passage. 
For comparison, we include the BT-SPLADE-L and uniCOIL model as baselines.

From Figure \ref{fig:splade-wo-expansion}, the uniCOIL (without expansion) model outperforms SPLADEv2 on 6/8 BEIR datasets, and BT-SPLADE-L on 4/8 BEIR datasets evaluated without including document expansion terms (the solid color portion in each graph), i.e. relying completely on the tokens present within the passage. However, factoring in the document expansion terms (the shaded portion in each graph), SPLADEv2 overall has the largest improvement from its own expansion terms in comparison to docT5query for the uniCOIL model. This provides a crucial insight about the behaviour of SPLADEv2, with the major boost in zero-shot performance gained on BEIR is due to expansion terms outside of the original passage.

\begin{table}[t]
    \centering   
    \setlength\tabcolsep{2.5pt}
    \begin{tabular}{l!{\color{lightgray}\vrule}l!{\color{lightgray}\vrule}r!{\color{lightgray}\vrule}r!{\color{lightgray}\vrule}r!{\color{lightgray}\vrule}r}
       
       \toprule

       \textbf{Model} &
       \textbf{Expansion}   &
       \multicolumn{1}{c!{\color{lightgray}\vrule}}{\textbf{DbPedia}}&
       \multicolumn{1}{c!{\color{lightgray}\vrule}}{\textbf{FiQA}} &
       \multicolumn{1}{c!{\color{lightgray}\vrule}}{\textbf{TREC-C.}} & \multicolumn{1}{c}{\textbf{Robust04}} \\

       \midrule
       \arrayrulecolor{lightgray}
       \multirow{2}{*}{\textbf{uniCOIL}} & docT5query & 0.361 & 0.298 & 0.710 & 0.422 \\
                               & TILDE & 0.382 & 0.294 & 0.637 & 0.424 \\
       \midrule
       \multirow{2}{*}{\textbf{TILDEv2}} & docT5query   & 0.321 & 0.266 & 0.716 & 0.382 \\
                                & TILDE        & 0.343 & 0.264 & 0.714 & 0.378 \\
       \midrule
       \multirow{2}{*}{\textbf{BM25}}  & docT5query  & 0.333 & 0.277 & 0.682 & 0.427 \\
                                & TILDE     & 0.306 & 0.226 & 0.664 & 0.405 \\
        \arrayrulecolor{black}
       \bottomrule
    \end{tabular}
    \vspace{1mm}
    \caption{Comparison of docT5query and TILDE document expansion techniques for uniCOIL, TILDEv2 and BM25 on four BEIR datasets: DbPedia, FiQA, TREC-COVID and Robust04.}
    \label{tab:tilde-vs-doct5query}
    \vspace{-8mm}
\end{table}

\RQthree
Recall that sparse models such as uniCOIL and DeepImpact rely upon external document expansion for expansion terms to enhance their vocabulary. 
From prior work, there are two popular approaches for passage expansion: DocT5query and TILDE.
DocT5query learns to generate synthetic queries for an input passage, whereas TILDE effectively learns to weight tokens from the BERT's vocabulary space, and the top 200 tokens with the highest weights are included as expansion terms. 
From Table \ref{tab:tilde-vs-doct5query}, we find that docT5query is a useful document expansion technique for BM25, as it severely outperforms TILDE. 
However, for both uniCOIL and TILDEv2 performance of TILDE and docT5query are mixed. 
Given the computational cost of docT5query is 140$\times$ more than TILDE, we find TILDE is a practical solution and keep it for future work to evaluate with all sparse retrieval systems.

\begin{table*}[t]
    \centering
    \small
    \vspace{-0.3cm}
    \begin{tabular}{p{1.8cm}|p{3.2cm}|p{3.6cm}|p{3.8cm}|p{3.6cm}}
    \toprule
    \textbf{Query (ID: 15)} & \textbf{Passage (ID: 325273)}	& \textbf{SPLADE V2 (BoW of Passage)} &	\textbf{with docT5query (q=20)}	& \textbf{with TILDE (k=200)} \\ \midrule
    Q: Can I send a \textcolor{red}{money order} from USPS as a business? & Doc: Sure you can.  You can fill in whatever you want in the From section of a \textcolor{red}{money order}, so your business name and address would be fine. The price only includes the \textcolor{red}{money order} itself.  You can hand deliver it yourself if you want, but if you want to mail it, you'll have to provide an envelope and a stamp... & \{ \textcolor{red}{order}: 2.1, \textcolor{red}{money}: 2.0, stamp: 1.7, mail: 1.7, receipt: 1.7, stu: 1.7, envelope: 1.6, section: 1.5, \textcolor{brown}{delivery}: 1.5, deliver: 1.4, fill: 1.4, address: 1.4, business: 1.4, \textcolor{blue}{fine}: 1.4, payment: 1.3, yourself: 1.3, record: 1.3, \textcolor{brown}{postal}: 1.3, stamps: 1.26, bank: 1.23, \textcolor{brown}{currency}: 1.21, contractor: 1.21, \textcolor{blue}{to}: 1.19, keep: 1.17, give: 1.14, records: 1.13, \textcolor{blue}{\#\#b}: 1.11, want: 1.1, delivered: 1.1, price: 1.1 ... \} & "Can you write a \textcolor{red}{money order} on a stub", "can i mail \textcolor{red}{money order} to a contractor", "how to send a \textcolor{red}{money order}", "can you mail \textcolor{red}{money order} yourself", "do i need a \textcolor{red}{money order} stamp", "can you hand deliver a \textcolor{red}{money order}", "how to send a \textcolor{red}{money order} without a stamp", "can someone hand deliver \textcolor{red}{money order}", "how long can someone deliver a \textcolor{red}{money order}" ... &    pay get orders cash need check cost much send letter copy take office paper receive  \textcolor{blue}{postalund} charge ordering use someone service long way deposit purchase house item instructions direct post transfer carry \textcolor{brown}{delivery} printed items needed number card paid buy sent us put done good sell company documents free required \textcolor{blue}{billco} form clerk \\ \bottomrule
    \end{tabular}
    \caption{Emperical analysis of document expansion techniques employed by various models on a sample query and its positive passage from FiQA-2018 dataset in the BEIR benchmark. The passage, SPLADEv2 (BoW of Passage) and docT5query generations have been overall truncated for demonstration purposes. }
    \label{tab:doc-exp-emperical-analysis}
    \vspace{-0.75cm}
\end{table*}

To understand the document expansion techniques better, we conduct an empirical analysis. 
Our visualizations are shown in Table \ref{tab:doc-exp-emperical-analysis}. Our findings are follows: (1) docT5query is found to hallucinate and repeat keywords from the passage such as ``money order'' (highlighted in red) across all its generated queries. 
This repetition assists lexical models as the keywords are multiplied and hence stronger keyword-based signals are generated, but does not add diversity, i.e. newer expansion terms into the original passage.
(2) TILDE only add keywords not present within the original and hence provides useful expansion terms, however, sometimes can misspell its expansion terms (highlighted in blue). 
(3) SPLADEv2 document expansion is overall viewed as useful, as it can weigh important keywords like ``money'' and ``expansion'' high. 
Further, it adds meaningful expansion terms such as ``postal'' or ``currency'' which are not present in the passage. 
SPLADEv2 can provide a high weight to nonsensical words such as ``\#\#b'', which highlights the fact that SPLADEv2 produces ``wacky weights'' \cite{wacky-weights}.

\begin{figure}[t]
\centering
    \includegraphics[trim=20 20 20 10, width=0.4\textwidth]{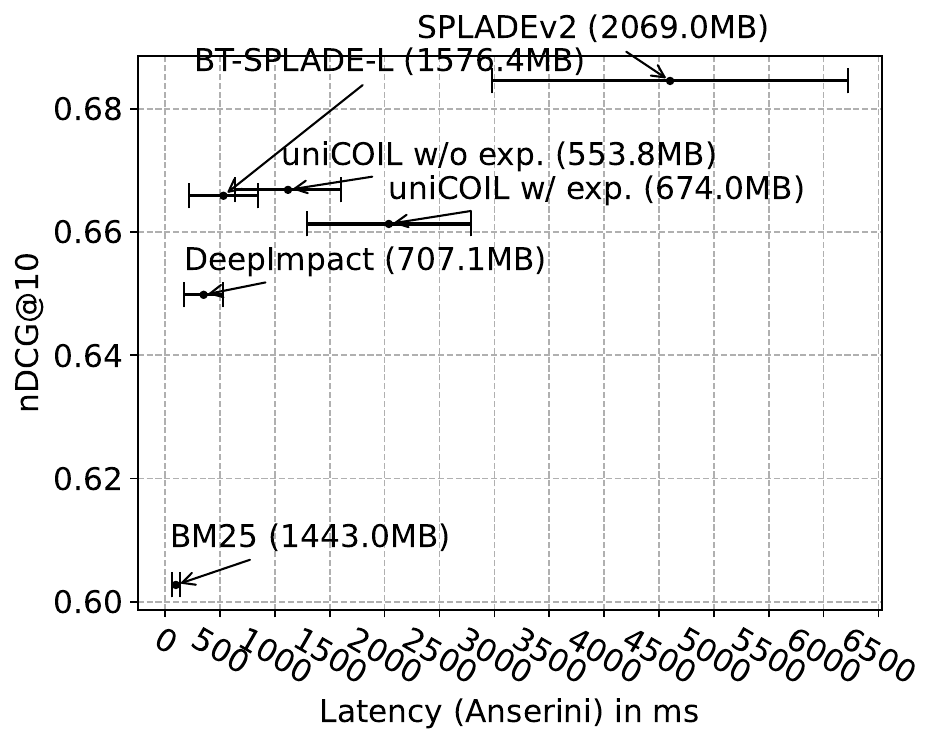}
    \caption{\textbf{Efficiency vs effectiveness on HotpotQA. Index sizes are marked within the parentheses.}}
    \label{fig:latency-vs-ndcg}
    \vspace{-3mm}
\end{figure}

\RQfour

\noindent
We are interested in evaluating efficiency of these sparse retrievers in terms of time and memory complexity. 
We use HotpotQA for this study, since it has a diverse range of query lengths from 7 words (counted with NLTK package) to 57 words. 
Following~\citet{wacky-weights}, we use one single thread on a CPU (Intel Xeon Platinum 8168 CPU @ 2.70GHz.) for efficiency evaluation. 

The pareto analysis is shown in \autoref{fig:latency-vs-ndcg}. Although sparse retrievers outperform BM25 on effectiveness (+0.047$\sim$0.071 nDCG@10), they are much slower (350ms$\sim$4600ms) than BM25 (97ms). 
Among all the models, BT-SPLADE-L achieves a good trade-off between latency (531ms) and effectiveness (nDCG@10 0.666), although it has a much larger index (1576.4MB), 2$\sim$3$\times$ as that of DeepImpact and uniCOIL. 
Interestingly from \autoref{fig:latency-vs-ndcg}, we find document expansion significantly increases the latency in retrieval. 
Adding document expansion terms slows down uniCOIL on average from 1119ms (w/o exp) to 2037ms (w/ exp) on HotpotQA. 

\begin{figure}[t]
\centering
    \includegraphics[trim=20 20 20 20, width=0.4\textwidth]{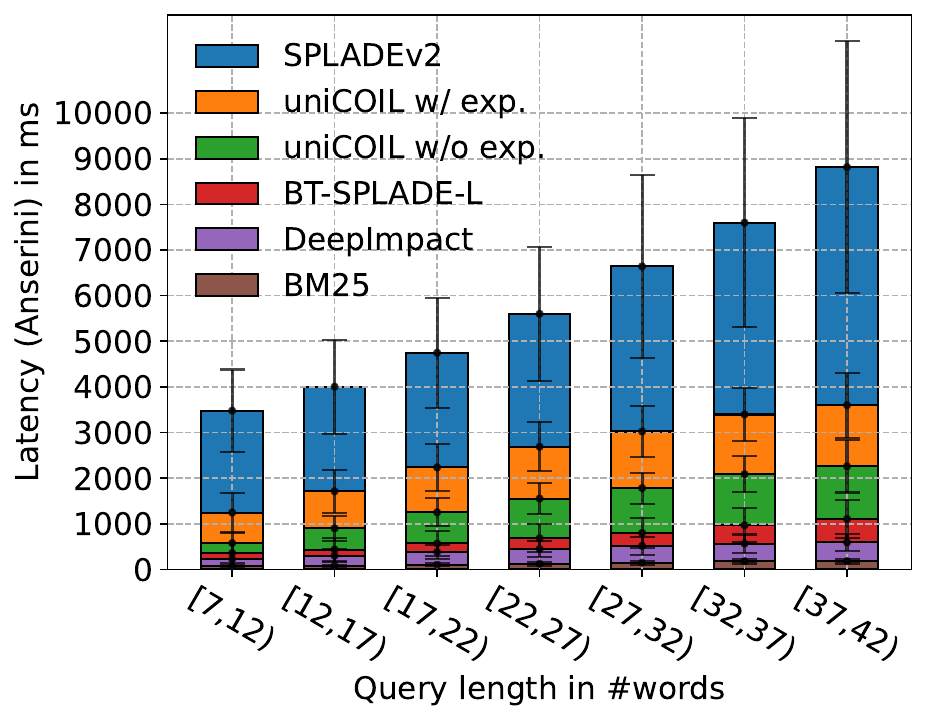}
    \caption{\textbf{Efficiency vs number of query length on HotpotQA.}}
    \label{fig:latency-vs-wlen}
\end{figure}

Together with~\autoref{tab:sparsity-analysis}, we find the query term sparsity influences latency noticeably more	 than the document term sparsity. 
To investigate this, we gather HotpotQA queries into bins of different lengths and report the average latency for each bin. The results are shown in~\autoref{fig:latency-vs-wlen}. 
Our results show that for all the sparse retrievers, both the mean values and the std. values of the latency increases linearly (all correlation coefficients > 0.99 on the binned data) with the query length. 
For example, BT-SPLADE-L latency increases from 358±221ms to 1105±417ms for [7, 12] to [37, 42) word length.

\section{Conclusion}

In this work, we presented \sprint, a unified Python toolkit focused on sparse neural retrieval. 
The toolkit extends the evaluation of several neural sparse retrievers on a common interface and easily allows practitioners to search their custom datasets using sparse retrieval. 
Evaluation of a custom dataset using our toolkit is straightforward, as we effectively use an inference pipeline to unify evaluation across all different sparse retrievers. 

With \sprint, we established strong zero-shot sparse baselines on BEIR. We found out that document expansion techniques are crucial and without (external) document expansion techniques, only SPLADEv2 and BT-SPLADE-L can significantly outperform the strong BM25 baseline. Next, we demystify zero-shot gains of SPLADEv2 on BEIR, and find major performance gained by the model is likely due to ``expansion terms'' outside of the passage.

\begin{acks}
This research was supported in part by the Natural Sciences and Engineering Research Council (NSERC) of Canada; computational resources were provided by Compute Canada. The work has been funded by the German Research Foundation (DFG) as part of the UKP-SQuARE project (grant GU 798/29-1). 
\end{acks}

\onecolumn
\begin{multicols}{2}
   \bibliographystyle{ACM-Reference-Format}
   \bibliography{camera_ready}
\end{multicols}



\end{document}